\documentclass[twocolumn,secnumarabic,amssymb, nobibnotes, aps, prb, superscriptaddress,longbibliography]{revtex4-2}

\usepackage{graphicx}
\usepackage{dcolumn}
\usepackage{bm}
\usepackage{lipsum}
\usepackage[dvipsnames]{xcolor}
\usepackage[utf8]{inputenc}
\usepackage[T1]{fontenc}
\usepackage{mathptmx}
\usepackage{etoolbox}
\usepackage{longtable}
\usepackage{float}
\usepackage{xcolor}
\usepackage{chemformula}
\usepackage{siunitx}
\usepackage{booktabs}

\begin{document}	
	\title{Statistical characterization of the spin Hall magnetoresistance in YIG/Pt heterostructures}
	\author{Denise Reustlen} \email{denise.reustlen@uni-konstanz.de}
	\affiliation{Department of Physics, University of Konstanz, 78457 Konstanz, Germany}
	
	\author{Sebastian Sailler} 
	\affiliation{Department of Physics, University of Konstanz, 78457 Konstanz, Germany}
	
	\author{Davina U. Schmidt}
	\affiliation{Department of Physics, University of Konstanz, 78457 Konstanz, Germany}

	\author{Richard Schlitz}
	\affiliation{Department of Physics, University of Konstanz, 78457 Konstanz, Germany}
	
	\author{Michaela Lammel} 
	\affiliation{Department of Physics, University of Konstanz, 78457 Konstanz, Germany}
	
	\author{Sebastian T. B. Goennenwein} \email{sebastian.goennenwein@uni-konstanz.de}
	\affiliation{Department of Physics, University of Konstanz, 78457 Konstanz, Germany}
	
	\date{\today}
	
	\begin{abstract}
            The spin Hall magnetoresistance (SMR) is widely used to study the interplay between charge and spin currents in bilayers of a magnetic insulator and a normal metal. However, not much is known about the spatial variation of the SMR across the surface of one and the same sample.
            In this work, we investigate the statistical distribution of the SMR in hundreds of nominally identical Hall bar structures patterned into prototypical yttrium iron garnet (YIG)/Pt heterostructures. We find a Gaussian-distributed SMR with a narrow standard deviation of approximately \SI{10}{\percent} of the mean value in each YIG/Pt bilayer studied.
            However, the variation of the mean SMR between different YIG/Pt samples can be as large as $\sim$\SI{30}{\percent}, despite nominally identical fabrication conditions.
            This demonstrates that spatial variations of the SMR amplitude must not be neglected, in particular when comparing different heterostructures. On a microscopic level, local variations of the interface quality captured by the spin mixing conductance are the most likely origin for the observed SMR amplitude variations. 
	\end{abstract}
	
	\maketitle
	Pure spin current phenomena enable the transfer of spin angular momentum without an accompanying charge flow, which makes them a backbone of modern spintronics \cite{sinova_spin_hall_effects_2015}. A quantitative understanding of how spin currents are generated, transmitted, and absorbed at interfaces is essential for optimizing spin transport efficiency and for advancing applications such as spin–orbit torque switching for magnetic memories \cite{Manchon_2019, Krizakova_2022}, spin Seebeck devices \cite{Uchida_2010, Bauer_2012, Kikkawa_2022}, and magnonic circuits \cite{Demidov_2020, Flebus_2024, Pirro_2021}. In this context, the spin Hall magnetoresistance (SMR) provides an effective probe of interfacial spin transfer processes. The SMR occurs in bilayers consisting of a magnetic layer (FM) and a normal metal (NM) layer \cite{nakayama_spin_2013, althammer_quantitative_2013, chen_theory_2013}. The SMR arises from a magnetization-orientation dependent spin transfer across the FM/NM interface \cite{nakayama_spin_2013, althammer_quantitative_2013, chen_theory_2013}, resulting in a change of the resistivity of the NM as a function of the magnetization orientation in the FM. The effect is quantified by the ratio of the magnetization-dependent resistivity change $\Delta \rho$ and the resistivity $\rho_\mathrm{0}$ of the NM itself: $\mathrm{SMR} = \Delta \rho / \rho_\mathrm{0}$. Typically, the NM consists of a heavy metal like Pt or Ta featuring a large spin Hall effect \cite{guo_intrinsic_2008, sinova_spin_hall_effects_2015, velez_hanle_2016}. Magnetic insulators like yttrium iron garnet (YIG) are the prototypical FM layer, as the charge current flow in the FM/NM bilayer is then limited to the NM layer.
 
 	After its initial discovery and theoretical description \cite{nakayama_spin_2013, chen_theory_2013} the SMR has been extensively studied \cite{Chen_2016}. 
        Next to the prototypical combination of YIG/Pt, the SMR has been observed in a manifold of different material combinations \cite{althammer_quantitative_2013, Ding_Fe3O4_2014, Aqeel_PRB_CoCr2O4_2015, isasa_spin_2016, kim_bilayers_2016, Hou_tunable_2017, Fischer_SMR_AFM_2018, leiviska2025spinhallmagnetoresistancealtermagnetic, velez_spin_2019, oyanagi_paramagnetic_2021}. 
        Between these materials, the value of the SMR can vary by multiple orders of magnitude with the largest values reported in YIG/Pt and \ch{Fe2O3}/Pt, being SMR = \SI{1.6e-3}{} and \SI{2.5e-3}{}, respectively \cite{althammer_quantitative_2013, Fischer_SMR_AFM_2018}. 
        While theory approaches modeling the SMR have been put forward, a quantitative analysis and comparison of the effect size for different material combinations remains to be performed \cite{zhang_theory_2019}. 
 	Such a modeling is further complicated, as even in YIG/Pt \cite{nakayama_spin_2013, althammer_quantitative_2013, marmion_temperature_2014, velez_competing_2016} the SMR effect amplitude differs significantly in the literature \cite{althammer_quantitative_2013, marmion_temperature_2014, velez_competing_2016}. 
        These differences can be traced back to multiple parameters determining the effect: the spin Hall properties of Pt \cite{sagasta_SHE}, the thickness of Pt \cite{althammer_quantitative_2013} and the interface quality defining the spin transparency \cite{velez_competing_2016,putter_impact_2017}. 
        While the Pt thickness and resistivity can be controlled via the deposition~\cite{althammer_quantitative_2013, sagasta_SHE}, its texture, crystallinity or grain size is more tedious to characterize \cite{sailler_2025}.
        Moreover, strong variations in the SMR amplitude can arise due to the interface quality \cite{Qiu_2013}.
        In particular, the YIG/Pt interface can either be obtained in an in-situ \cite{althammer_quantitative_2013} or ex-situ process \cite{jungfleisch_improvement_2013, putter_impact_2017}.
        The largest SMR in YIG/Pt is observed for in-situ interfaces~\cite{althammer_quantitative_2013}, whereas the initial surface treatment of the YIG layer largely impacts the resulting SMR effect size in an ex-situ process \cite{jungfleisch_improvement_2013, Vlietstra_2013, velez_competing_2016, putter_impact_2017, Li_2020}. 
 	
 	Despite the large spread in SMR amplitudes and parameter values, the SMR amplitude for a given FM/NM bilayer sample is commonly extracted from measurements on a single Hall bar device and taken as a representative value for this material combination.
    Interestingly, not much is known about the reproducibility and the statistical error across the surface of a single sample, not to speak of series of bilayer samples fabricated using a nominally identical fabrication procedure.

 	In this work, we compare the SMR response of three nominally identical YIG/Pt bilayer samples by systematically studying hundreds of Hall bar devices patterned onto each sample. Our experiments yield insights into the statistical behavior of the SMR and its spatial homogeneity. 
        We find a Gaussian distribution of the SMR on each sample with a narrow standard deviation of about \SI{10}{\percent}, whereas between the samples the SMR can vary by up to \SI{30}{\percent}. 
        
        We utilize commercially available YIG films epitaxially grown via liquid phase epitaxy as the magnetic layer. 
        Before the Pt deposition, the surface is cleaned with Piranha-acid (\ch{H2O2}:\ch{H2SO4}) as described in Ref.~\cite{Schlitz_interface}.
        A \SI{5}{nm} Pt layer is then deposited via magnetron sputtering in an ultra high vacuum deposition system with a base pressure smaller than \SI{1e-7}{\milli\bar}, at a RF power of \SI{50}{\watt} yielding a deposition rate of \SI{1.8}{\nano\meter\per\minute}. In total we prepared three samples, referred to as S1, S2 and S3.
	After deposition, Hall bars are defined using optical lithography and Ar$^+$ ion etching. 
    
\begin{figure}[t]
		\begin{center}
			\includegraphics[width=\linewidth]{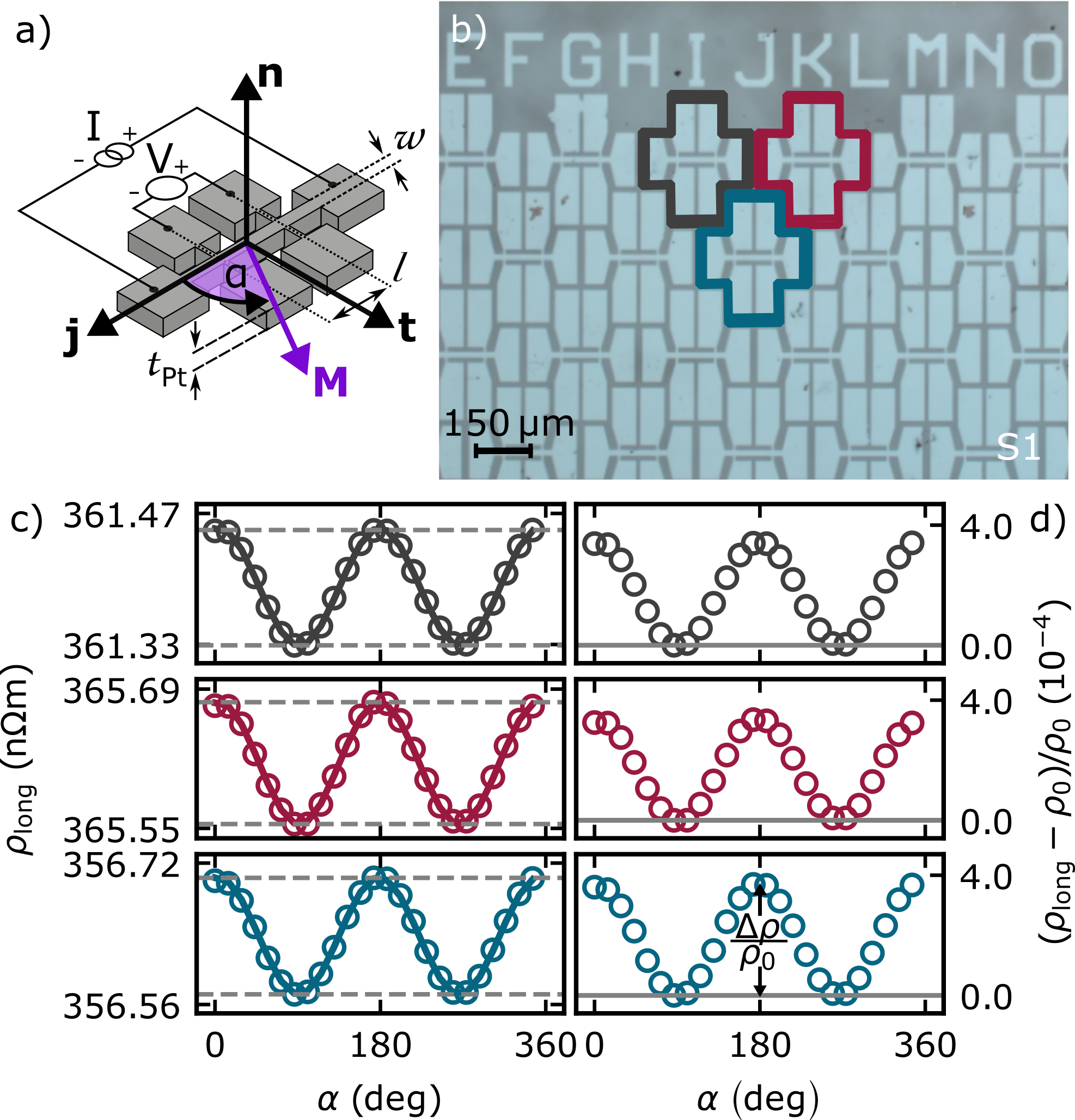}
                        \caption{a) Hall bar device and coordinate system used throughout this study. Via magnetron sputtering, a Pt layer being $t_\mathrm{Pt}$ thick is deposited onto a given YIG film. By using optical lithography hundreds of Hall bars with width $w$ and length $l$ are patterned into the Pt layer. A current $I$ is driven through the device and the voltage $V$ is measured. The magnetization of the YIG layer is rotated in the film plane by an external magnetic field applied at a varying angle $\alpha$. b) Micrograph of sample S1. c) Resistivity modulation and d) normalized magnetoresistance obtained from the three framed Hall bars in b). The SMR ($\Delta \rho/\rho_\mathrm{0}$) is indicated for the measurement shown in blue.}
			\label{Fig_1}
		\end{center}
	\end{figure}

    Figure~\ref{Fig_1}a) schematically depicts a Hall bar structure. Figure~\ref{Fig_1}b) shows a micrograph of the patterned sample.
        The magnetoresistance response of the Hall bars are characterized in a custom build probe station.
        A rotatable permanent magnet Halbach array is used to rotate a magnetic field of constant magnitude within the YIG/Pt interface plane, and thus to control the in-plane magnetization of the YIG layer. The angle between the current direction $\textbf{j}$ and the magnetization $\textbf{M}$ is defined as $\alpha$.
        Micropositioners with tungsten tips are used to contact a given Hall bar to apply a current $I$ along the Hall bar and to measure the ensuing voltage $V$. As the sample is mounted onto an x-y-z stage, our probe station allows measuring the SMR in multiple Hall bars in a fully automated fashion. By taking the Hall bar dimensions into account the resistivity $\rho_\mathrm{long}$ and the field-orientation dependent resistivity change $\Delta \rho$ can be calculated.\newline
        Figure~\ref{Fig_1}c) exemplarily shows the SMR observed in three different Hall bars framed in Fig.~\ref{Fig_1}b). 
        The SMR manifests as a $\cos^2(\alpha)$ modulation of the longitudinal resistivity \cite{althammer_quantitative_2013, nakayama_spin_2013}.
        We fit the data by using 
        \begin{align}
        \label{eq:SMR}
        \rho_\mathrm{long} = \Delta\rho\cdot\cos^2{(\alpha+\alpha_\mathrm{0})}+\rho_\mathrm{0},
        \end{align}
        and calculate the SMR amplitude as $\mathrm{SMR}=\Delta \rho/\rho_\mathrm{0}$. Here, $\alpha_\mathrm{0}$ corresponds to a phase shift of the SMR modulation, which can occur due to a misalignment of the external field with respect to the sample in the measurement setup. The SMR in the three Hall bars is SMR = \SI{3.48e-4}{}, \SI{3.37e-4}{} and \SI{3.72e-4}{}, respectively. This emphasizes a similar SMR effect in all measurements. Small deviations of $\Delta\rho$ and $\rho_\mathrm{0}$ on the scale of \SI{10}{\percent} and \SI{3}{\percent} are noticeable between the different devices in Fig.~\ref{Fig_1}c) and d). Possible causes are discussed in the following.
    \begin{figure}[t]
		\begin{center}
			\includegraphics[width=\linewidth]{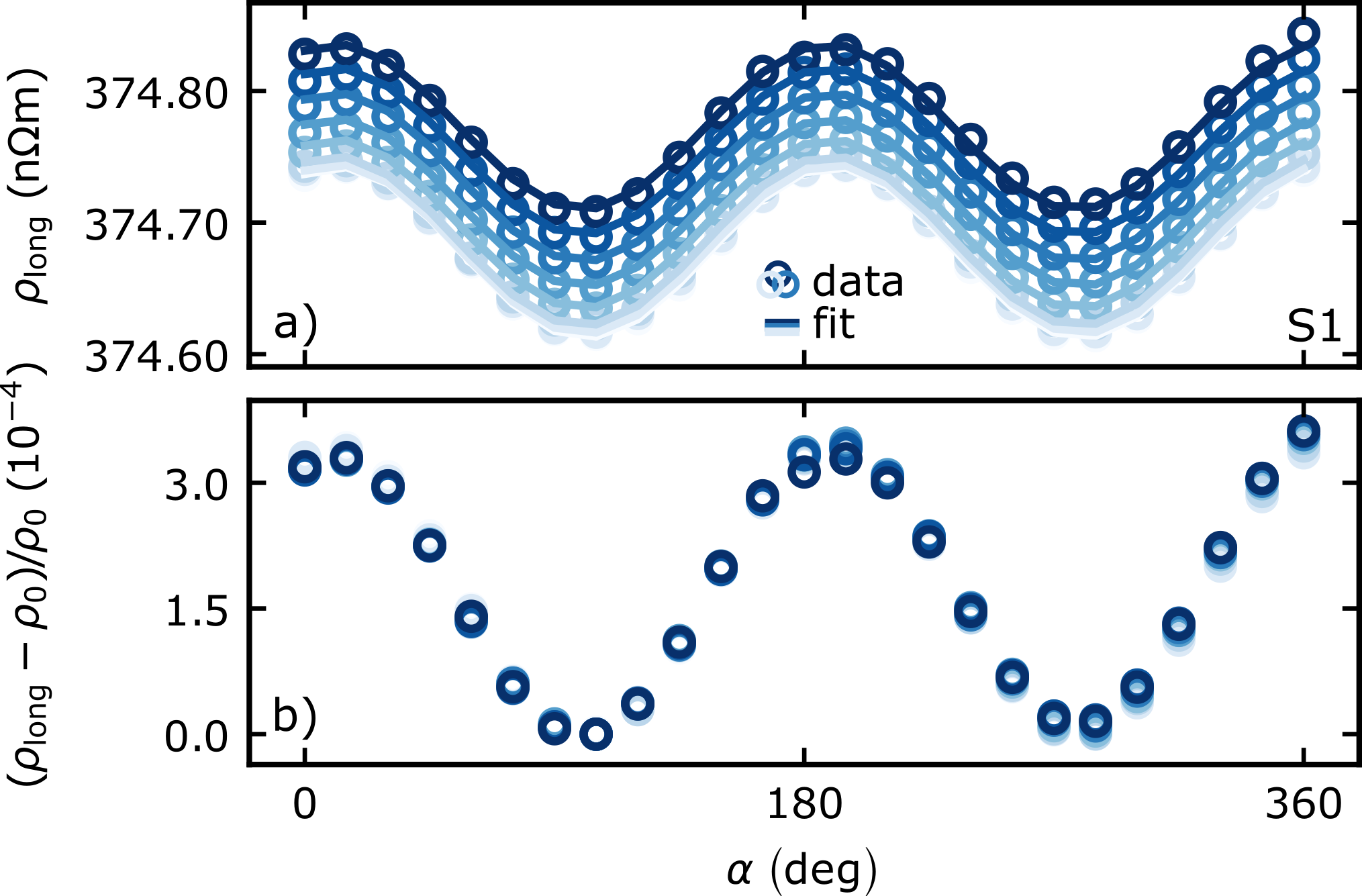}
                        \caption{a) Repeatability of the measurement scheme on a single Hall bar. The contacting and measuring scheme was performed 8 times over an hour. Due to slight laboratory temperature changes, the resistivity of the Pt layer changes. The total temperature change corresponds to \SI{0.15}{\kelvin}. The fit to the data yields $\alpha_\mathrm{0} =$ \SI{-11.1}{\degree}, which accounts for the misalignment of the external field. b) The normalized data for all performed measurements yield a mean SMR of \SI{3.41e-4}{}, with the standard deviation $\sigma=$ \SI{0.04e-4}{}, emphasizing the reproducibility of our measuring scheme.}
			\label{Fig_2}
		\end{center}
	\end{figure}
    
        In a first iteration, we discuss the influence of the external temperature on the measurement results and the extracted SMR. The setup is not temperature stabilized and the Pt resistivity is very sensitive to changes of the temperature. This influence is highlighted by the Pt temperature coefficient of \SI{5.752e-3}{\per\kelvin} (determined from a calibration measurement). Thus, changes in resistivity can stem from temperature variations during the measurement of the devices. 
        
        To assert that the normalization of $\Delta\rho$ to $\rho_\mathrm{0}$ does not hide additional effects and gauge the reproducibility of the automated contacting, we perform several measurements over the course of an hour on one Hall bar.
        Figure~\ref{Fig_2}a) depicts the resistivity of 8 measurements performed on the same Hall bar. The resistivity offset $\rho_\mathrm{0}$ varies significantly between the measurements. Using the temperature coefficient introduced earlier, the change in resistivity can be explained by a temperature change of \SI{0.15}{\kelvin}. Thus, the differences in resistivity can be attributed to changes in the external temperature. After normalization of the data, very similar SMR amplitudes become apparent with a narrow standard deviation of $\sigma$ = \SI{4e-6}{} which is less than \SI{2}{\percent} of the SMR amplitude (cp. Fig.~\ref{Fig_2}b). This result demonstrates that our measurement technique yields a robust SMR value on each device and is suited for the statistical investigation of the SMR. 
    
\begin{figure}[t]
		\begin{center}
			\includegraphics[width=\linewidth]{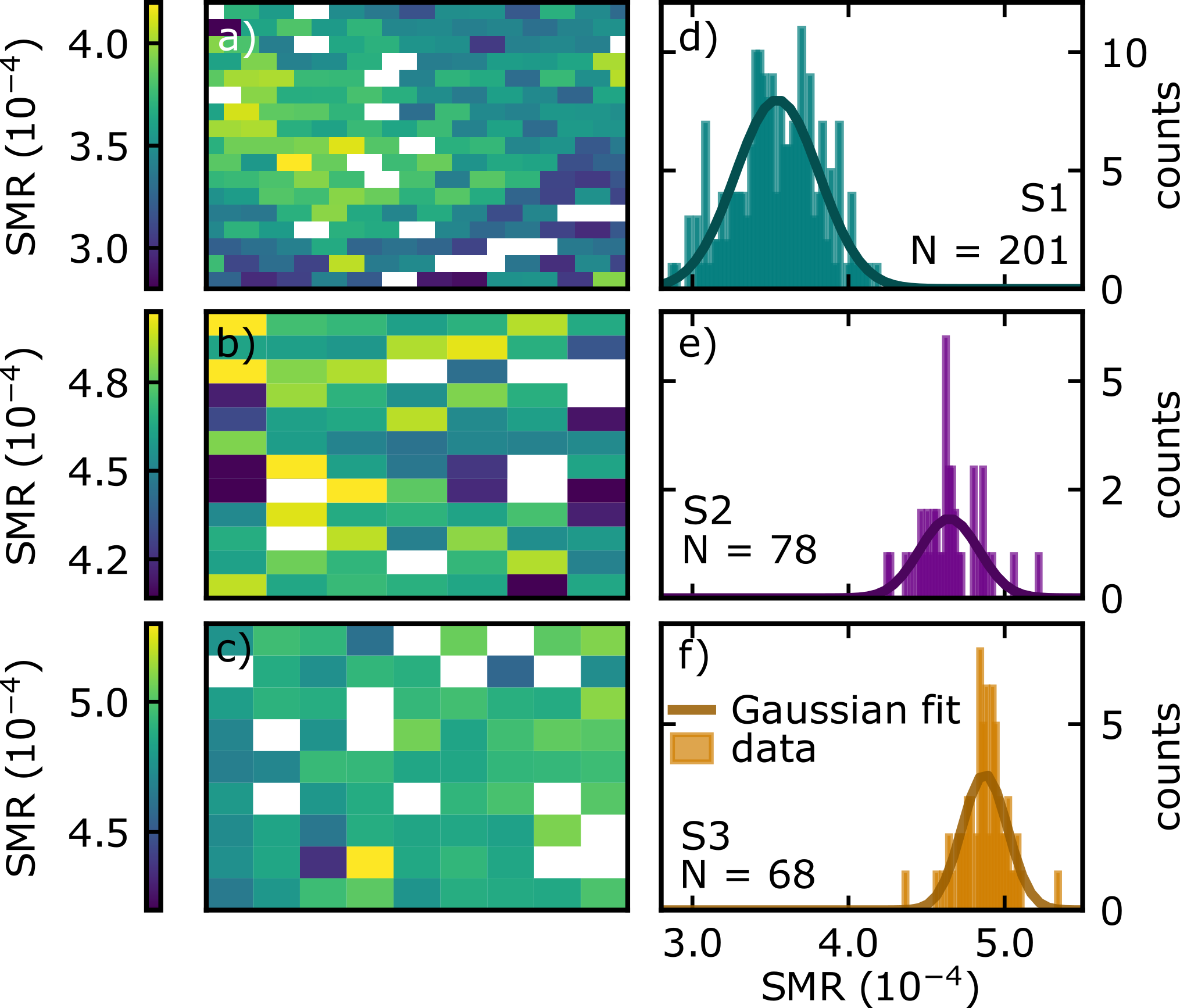}
                        \caption{a), b) and c) Color maps of S1, S2 and S3. Every rectangle corresponds to a Hall bar, where the color of the individual rectangle represents the value of the SMR: In all color maps, the SMR is quite homogeneously distributed, however, a small scattering of the SMR values is visible. Geometrical factors stemming from the fabrication process of the Hall bars can be neglected because of the normalization. The temperature can also be excluded as an explanation, as the variation could only be explained by assuming large temperature changes. Ultimately, intrinsic effects impacting the SMR such as the spin transparency of the YIG/Pt-interface are left in order to explain the standard deviation. d), e) and f) Histograms of the SMR amplitude distribution for samples S1, S2 and S3. For all three samples, a Gaussian distribution of the SMR, can be observed, yielding mean values $\mu$ of \SI{3.54e-4}{}, \SI{4.61e-4}{} and \SI{4.87e-4}{} and standard deviations $\sigma$ of \SI{0.27e-4}{}, \SI{0.23e-4}{} and \SI{0.14e-4}{}, respectively. Therefore, the variation of the SMR on one sample is as small as \SI{10}{\percent}. However, the SMR variation across multiple samples can be up to \SI{30}{\percent}. }
			\label{Fig_3}
		\end{center}
	\end{figure}

    We thus turn to the statistical analysis of the SMR by investigating sample S1 with N = 225 Hall bars.
    We measure the resistivity of all Hall bars and extract their magnetization orientation dependent resistivity change and resistivity offset. Such a large set of measurements allows to determine the statistical distribution of the SMR. 
    In Fig.~\ref{Fig_3}a), the spatial distribution of the SMR is depicted. Each rectangle represents an individual Hall bar device, with the positions in the color map corresponding to the position on the sample. White spaces describe shorted or broken devices.
    
        In total more than \SI{90}{\percent} of the devices functioned and are evaluated.
	A consistently large amplitude around SMR = \SI{4e-4}{} is found across the whole sample. However, a slightly brighter area, corresponding to a larger SMR amplitude can be observed on the left. The SMR can be expressed via Eq. \eqref{eq:SMR1}\cite{chen_theory_2013}:
        \begin{align}
            \label{eq:SMR1}
            \mathrm{SMR} = \frac{2\theta_\mathrm{SH}^2\lambda^2(\frac{\rho_\mathrm{0}}{t_\mathrm{Pt}})G\big(\tanh^2(\frac{t_\mathrm{Pt}}{2\lambda})\big)}{1+2\lambda\rho_\mathrm{0}G\bigg(\big(\tanh(\frac{t_\mathrm{Pt}}{\lambda})\big)^{-1}\bigg)}.
        \end{align}
    Here, $\theta_\mathrm{SH} = \rho_\mathrm0\cdot\sigma_\mathrm{SH}$ is the spin Hall angle with the spin conductivity $\sigma_\mathrm{SH}$ and $\lambda = \sigma_\mathrm{Pt}\cdot c$ is the spin diffusion length with $\sigma_\mathrm{Pt}=\frac{1}{\rho_\mathrm{0}}$ and $c =$ \SI{0.61e-15}{\ohm\square\meter} \cite{sagasta_SHE, sailler_2025}. Equation \eqref{eq:SMR1} implies that the SMR depends on many parameters. Especially the spin mixing conductance $G$ is demanding to control in the fabrication process. Therefore, variations of the SMR across the surface of a given bilayer sample are most likely to stem from a change of $G$ across the sample.
    
     Figure~\ref{Fig_3}d) depicts a histogram of the extracted SMR ($\Delta \rho / \rho_\mathrm{0}$), where similar SMR values are binned. Shorted or broken devices are excluded. The histogram reveals a Gaussian distribution of the SMR across the sample. This information is crucial to establish how deviations of the SMR ratio can be associated to variations between devices on a sample. To statistically analyze the distribution of the SMR, the data is fit by: 
        \begin{equation}
                f(\mathrm{SMR}) = \frac{1}{\sigma \times 2\pi} e ^{-\frac{1}{2}\bigg(\frac{\mathrm{SMR} - \mu}{\sigma}\bigg)^2},
		  \label{eq_gauss}
	    \end{equation}
        where $\mu$ is the mean value and $\sigma$ the standard deviation. We find $\mu_\mathrm{S1}$ = \SI{3.54e-4}{} and $\sigma_\mathrm{S1}$ = \SI{0.27e-4}{}, see Tab.~\ref{tab_1}, where $\sigma_\mathrm{S1}$ is well beyond the measurement reproducibility.
        \begin{table}[b]
        \caption{Mean value $\mu_\mathrm{SMR}$ and standard deviation $\sigma_\mathrm{SMR}$ of the SMR, as well as the mean resistivity $\mu_{\mathrm{\rho_\mathrm{0}}}$ and corresponding standard deviation, for the three samples investigated.}
        \label{tab_1}
        \begin{tabular}{c|c|c|c|c}
        \hline\hline
        samples & $\mu_{\mathrm{SMR}}$ & $\sigma_{\mathrm{SMR}}$ & $\mu_{\rho_0}$ (n$\Omega$m) & $\sigma_{\rho_0}$ (n$\Omega$m) \\ \hline
        S1 & $3.54 \times 10^{-4}$ & $0.27\times ^{-4}$ & 366 & 10.9 \\ \hline
        S2 & $4.61 \times 10^{-4}$ & $0.23 \times 10^{-4}$ & 283 & 6.7 \\ \hline
        S3 & $4.87 \times 10^{-4}$ & $0.14 \times 10^{-4}$ & 449 & 11.5 \\ 
        \hline\hline
        \end{tabular}    
        \end{table} 

    Repeating this statistical assessment of the SMR amplitude for the other two nominally identical samples (S2 and S3), the SMR can be compared between different YIG/Pt heterostructures. To that end, the SMR is measured and extracted as previously described, and exhibits a very similar behavior to sample S1, as shown in in Figs.~\ref{Fig_3}b) and c). Figures~\ref{Fig_3}e) and f) depict the corresponding SMR distribution across the respective sample. The Gaussian distribution in Figs.~\ref{Fig_3}e) and f) correspond to $\mu_\mathrm{S2}$ = \SI{4.61e-4}{} and $\mu_\mathrm{S3}$ = \SI{4.87e-4}{}, the standard deviations are summarized in Tab.~\ref{tab_1}. The histograms of all three samples (S1-S3) can be consistently described via a Gaussian distribution with a narrow standard deviation [see Fig.~\ref{Fig_3}d)-f)]. Overall, a robust value of the SMR is observed on each sample. However, the mean SMR value between these samples varies significantly.

	The mean SMR values of S1, S2 and S3 are in good agreement with the literature. Similar SMR values of roughly \SI{3e-4}{} were also obtained by using a similar surface treatment as in this study \cite{putter_impact_2017}. The observed SMR amplitudes furthermore fall between those reported for in-situ interface (\SI{1e-3}{}, \cite{althammer_quantitative_2013}) and samples prepared without prior surface treatment \cite{velez_competing_2016}. This comparison shows deviations well beyond the observed standard deviation, suggesting that despite nominally identical conditions, a significant variation of the observed SMR is present. 

	The standard deviations summarized in Tab.~\ref{tab_1} signalize a narrow distribution and a robust value across the individual samples. 
        Consequently, on hindsight, the measurement of a single device on a YIG/Pt sample provides a reasonable representation of the sample quality and the actual size of the SMR. However, as revealed by our statistical analysis, the precise magnitude of the SMR amplitude can only be estimated from the response of one single Hall bar with an uncertainty of (several) \SI{10}{\percent} \cite{althammer_quantitative_2013,putter_impact_2017, Chen_2016}.\newline
    \begin{figure}[t]
		\begin{center}
			\includegraphics[width=\linewidth]{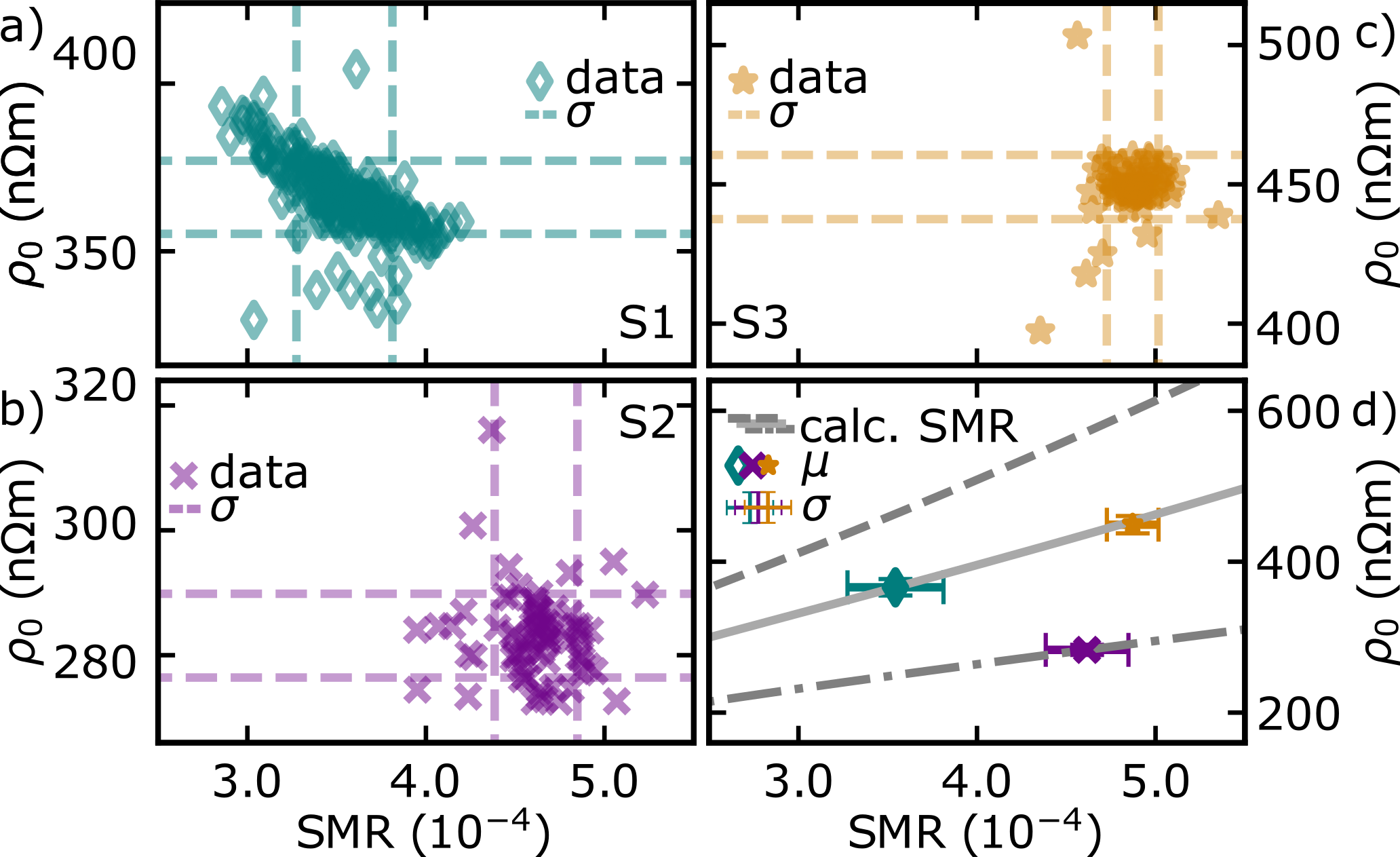}
                        \caption{a), b) and c) Correlation of the SMR and the resistivity $\rho_\mathrm{0}$ for every Hall bar measured on S1, S2 and S3. The dashed lines indicate the standard deviation of the SMR amplitude and $\rho_\mathrm{0}$, respectively. In a), if the SMR values increase, the resistivity decreases and vice versa, whereas the opposite behavior is expected from Eq.~\eqref{eq:SMR1}. In b) and c), this behavior is not reproduced suggesting that the observed trending is unique to sample S1. d) The mean of all SMR values and resistivity values per sample. The error bars indicate the standard deviation of both the SMR and $\rho_\mathrm{0}$. Additionally, the gray line is a calculation of the SMR according to Eq.~\eqref{eq:SMR1}, revealing a trending depending on $\rho_\mathrm{0}$, which reproduces the measured data well. We used $G=$\SI{3e14}{\ohm^{-1}\metre^{-2}} for the dashed line, $G=$\SI{5e14}{\ohm^{-1}\metre^{-2}} for the solid line, and $G=$\SI{3e15}{\ohm^{-1}\metre^{-2}} for the dash-dotted line.}
			\label{Fig_4}
		\end{center}
	\end{figure}
        We now discuss the variation of the SMR in a single and between different samples and further its dependency on the resistivity. To that end, Fig.~\ref{Fig_4} summarizes the SMR versus the corresponding $\rho_\mathrm{0}$ for every Hall bar of all three samples S1, S2 and S3 [see Figs.~\ref{Fig_4} a) - c)]. The dashed lines indicate the standard deviation of the SMR and of the resistivity (see also Tab.~\ref{tab_1}). 

        The Hall bar geometry, temperature as well as the YIG/Pt interface quality could influence the correlation of the SMR and the resistivity. As already observed in Fig.~\ref{Fig_2}, the temperature strongly influences the resistivity $\rho_\mathrm{0}$. By assuming that the temperature does not change more than \SI{5}{\kelvin} during the entire measurement run, only a resistivity change of \SI{3.19}{\nano\ohm\meter} can be explained. However, the standard deviation clearly exceeds this by a factor of 3.4. Additionally, a minor impact of the temperature on the SMR is well established \cite{Meyer_2014} and is assumed to behave linearly in a range from \SI{150}{\kelvin} to \SI{300}{\kelvin}. A temperature change of \SI{5}{\kelvin} leads to a SMR change of \SI{0.23e-05}{}, which is almost an order of magnitude smaller than the standard deviations of all measured samples. Ultimately, laboratory temperature variations thus cannot account solely for the observed variations of $\rho_\mathrm{0}$ and the SMR across the sample surface.
        
        In order to exclude the variations of the Hall bar geometries, a few Hall bars were randomly chosen and their width and length measured, yielding a variation of up to \SI{5}{\percent} in both dimensions. In turn, the thickness $t_\mathrm{Pt}$ of the Pt layer on S1 was analyzed via ellipsometry on 6 different locations, all resulting in $t_\mathrm{Pt}$ = \SI{5.1}{\nano\meter} with variations not larger than \SI{0.3}{\nano\meter}. We thus conclude that the variations in the actual Hall bar geometry are the likely origin of the observed standard deviation in $\rho_\mathrm{0}$. However, since the geometry does not enter into the SMR due to the normalization of $\Delta \rho$ to $\rho_\mathrm{0}$, geometrical variations cannot explain the observed deviations of the SMR.
        
        Having excluded extrinsic effects, the observed variation of the SMR evident from Fig.~\ref{Fig_4} must stem from intrinsic properties of the YIG/Pt heterostructure. In the following, we discuss potential changes of the intrinsic spin Hall properties of the Pt film. Here, the spin Hall angle $\theta_\mathrm{SH}$ and the spin diffusion length $\lambda$ depend on $\rho_\mathrm{0}$ \cite{sagasta_SHE}. Large $\rho_\mathrm{0}$ correspond to large $\theta_\mathrm{SH}$, whereas large $\rho_\mathrm{0}$ correspond to small $\lambda$. Both parameters influence the SMR [see Eq.~\eqref{eq:SMR1}]. However, in Fig.~\ref{Fig_4} a), large SMR values systematically correspond to small resistivity values and vice versa. Neither thickness variations nor the scaling of $\theta_\mathrm{SH}$ and $\lambda$ explain this non trivial behavior. We interpret this as evidence that $\theta_\mathrm{SH}$ and lambda are not the main cause for the variations in the SMR amplitude across the sample surface. 
        
        This leaves the YIG/Pt interface quality as a possible cause for the scaling. Generally, the spin mixing conductance $G$ seems to be homogeneous across the sample, however small deviations cannot be excluded. The trending of the correlation in Fig.~\ref{Fig_4}a) seems to be uniquely attributed to this sample and is absent in the other samples [see Fig.~\ref{Fig_4}b) and c)]. The origin of the correlation cannot be pinpointed further.
        
    In Fig.~\ref{Fig_4} d), the mean of all SMR and resistivity values as well as their corresponding standard deviations is plotted. As stated before, large $\rho_\mathrm{0}$ should be accompanied by large SMR values [cp. Eq.~\eqref{eq:SMR1}]. Three different spin mixing conductances $G=$ \SI{5e14}{\ohm^{-1}\metre^{-2}} (solid line), $G=$ \SI{3e14}{\ohm^{-1}\metre^{-2}} (dashed line) and $G=$ \SI{3e15}{\ohm^{-1}\metre^{-2}} (dash-dotted line) are used to model the SMR as a function of $\rho_\mathrm{0}$ (see Eq. \eqref{eq:SMR1}). This scaling is depicted in Fig.~\ref{Fig_4} d). Here, for $G=$ \SI{5e14}{\ohm^{-1}\metre^{-2}}, the SMR model resembles the mean SMR obtained in sample S1 (green) and S3 (yellow) well. However, in S2 (violet) a somewhat different spin mixing conductance of $G=$ \SI{3e15}{\ohm^{-1}\metre^{-2}} matches the data best.
    
    The spin mixing conductance is the only parameter which cannot be controlled easily across the entire sample in this study. Therefore, we conclude that both the scattering of the SMR on an individual sample as well as the deviations across several samples are most likely attributed to a varying spin mixing conductance between the individual samples and across their surfaces, respectively. 
    Despite these variations, we overall can reproduce a consistent SMR value across the sample with a narrow standard distribution. 
    
    In summary, we have investigated the SMR in multiple YIG/Pt bilayer samples. Onto each bilayer, we patterned numerous Hall bar devices and investigated the statistics of the SMR. We find a mean value $\mu$ = \SI{3.54e-4}{} and a standard deviation $\sigma$ = \SI{0.27e-4}{} for sample S1 with 225 devices. The standard deviation is approximately \SI{10}{\percent} of $\mu$ which is reproduced in the other two, nominally identical samples. We conclude that the SMR is Gaussian distributed and reproducible across a given sample, but can vary significantly between samples. The dominant source of the scattering of the SMR is found to be the spin mixing conductance, leading to variations of $\sim$ \SI{30}{\percent} of the SMR when comparing different samples. Our results show that spatial variations of the spin mixing conductance must be carefully considered when investigating small variations in the SMR, especially between different samples.

    This work was funded by the Deutsche Forschungsgemeinschaft (DFG, German Research Foundation) via the SFB 1432, Project-ID No. 425217212 and via Project-ID No. 490730630. We also gratefully acknowledge discussions with M. Kuepferling, P. Moehrke and M. Hagner, as well as technical support via the nano.lab core facility of the University of Konstanz.
    
	\bibliography{Reustlen_SMR_statistics.bib}
	
\end{document}